\documentclass[11pt]{llncs}
\usepackage{graphicx}
\usepackage{cite}
\usepackage{url}
\usepackage{listings}
\newenvironment{listing}{\table}{\endtable}
\newenvironment{listing*}{\table*}{\endtable*}

\DeclareSymbolFont{AMSb}{U}{msb}{m}{n}
\DeclareSymbolFontAlphabet{\Bbb}{AMSb}

\makeatletter
\def\hb@xt@{\hbox to }
\makeatother

\let\oldendproof\endproof
\def\endproof{\qed\oldendproof}

\begin{document}

\title{All Maximal Independent Sets and\\ Dynamic Dominance for Sparse Graphs} 

\author{David Eppstein\thanks{Supported in part by NSF grant CCR-9912338.}}

\institute{Computer Science Department\\
School of Information \& Computer Science\\
University of California, Irvine\\
\email{eppstein@uci.edu}}

\maketitle   

\begin{abstract}
We describe algorithms, based on Avis and Fukuda's reverse search paradigm, for listing all maximal independent sets in a sparse graph in polynomial time and delay per output.
For bounded degree graphs, our algorithms take constant time per set generated;
for minor-closed graph families, the time is $O(n)$ per set, and for more general sparse
graph families we achieve subquadratic time per set.
We also describe new data structures for maintaining a dynamic vertex set $S$ in a sparse or minor-closed graph family, and querying the number of vertices not dominated by $S$; for minor-closed graph families the time per update is constant, while it is sublinear for any sparse graph family.  We can also maintain a dynamic vertex set in an arbitrary $m$-edge graph and test the independence of the maintained set in time $O(\sqrt m)$ per update.  We use the domination data structures as part of our enumeration algorithms.
\end{abstract}

\section{Introduction}

In this paper we revisit the classical combinatorial enumeration problem of generating all maximal independent sets of a graph, or equivalently generating all cliques in the complement of the graph.
Many algorithms for this problem are known; see, e.g.,~\cite{JenMot-LATIN-92,JohYanPap-IPL-88,LawLenRin-80,Leu-Algs-84,LiaDhaLak-SAC-91,MakUno-SWAT-04,MisPit-COLT-97,Sti-COA-04,TsuIdeAri-SJC-77,YuChe-IJCM-93}, or the survey by Bomze et al.~\cite{BorBudPar-HCO-99}. However, all take at least quadratic time per generated independent set even on sparse graphs. There has also been work on maximal independent set enumeration with time bounds proportional
to the maximum possible number of such sets~\cite{Bys-SODA-03,Epp-JGAA-03,TomTanTak-COCOON-04}, but these algorithms may take a large time per set on instances with few maximal independent sets.
We find more efficient maximal independent set generation algorithms for many graph classes including bounded degree graphs, minor-closed graph families, and subgraph-closed sparse graph families. Although graphs in these classes have polynomially many cliques~\cite{ChiNis-SJC-85}, generating all maximal independent sets is more difficult for these graphs as there may be exponentially many such sets.

Our maximal independent set algorithms require us to quickly test whether certain subsets of the vertices of the graph dominate the remaining vertices, so we also consider the problem of dynamic dominance testing.  We provide very efficient algorithms for this problem on minor-closed graph families, taking constant time per test.  We also find sublinear algorithms for dynamic dominance in more general sparse graph families.  As a simple consequence of the latter result we show how to test independence of dynamic vertex sets in arbitrary graphs.

\section{Reverse Search}

{\em Reverse search} is a powerful paradigm for enumeration developed by Avis and Fukuda~\cite{AviFuk-DCG-92,AviFuk-DAM-96} and applied by them to many enumeration problems including listing all  vertices of a convex polytope, triangulations of a planar point set, vertices or cells of a hyperplane arrangement, spanning trees in graphs and non-crossing spanning trees in the plane, connected induced subgraphs of graphs, and topological orderings of directed acyclic graphs.

To use reverse search, one needs a {\em parent} operation on the objects being enumerated, such that any two objects can be transformed to a common {\em canonical object} by repeated parent operations.  For instance, for triangulations, the parent operation finds two triangles sharing an edge and maximizing
the sum of the two angles opposite the shared edge, and replaces them by two other triangles with the same union; repeated application of this flip operation reduces any triangulation to a canonical triangulation, the Delaunay triangulation of the input.  For convex polytopes, the parent operation can be any simplex method pivot rule, for an appropriate linear objective function, and the canonical object is the vertex minimizing the objective function's value.
We form a rooted tree, having as its vertices the objects to be enumerated, with edges between each object and its parent, and having the canonical object as root.  Reverse search is simply depth first tree traversal applied to this tree.  In order to perform this traversal, one must be able to identify the children of each node in the tree.  Typically, a polynomially-sized superset of the children is identified (e.g. in the case of triangulations, this superset could be all triangulations reachable from the given one by a single flip).  We can then test whether any member of the superset is a child of the current object by applying the parent operation to it and comparing the result to the current object.

Thus, reverse search performs the following steps for each object: identify a superset of the object's children, apply the parent operation to test each member of the superset, and search recursively each child found by this test.  In this basic form, the time per object is the product of the number of items in the superset with the time to generate each member of the superset and apply the parent operation to it.  As we shall see for our maximal independent set enumeration algorithm, however, reverse search may be sped up in various ways, for instance by maintaining data structures that allow us to perform the parent tests more quickly.

\section{Basic Independent Set Generation Algorithm}

\def\LFMIS{{\rm LFMIS}}
\def\parent{\mathop{\rm parent}}
\def\later{\mathop{\rm later}}

To apply reverse search to maximal independent set generation, we need an appropriate parent for each maximal independent set~$S$.  Assume we have ordered the vertices into a sequence; this allows us to compute the lexicographically first maximal independent superset of any independent set by considering the vertices one at a time, in order, and adding each vertex to the set whenever it is independent of the vertices already in the set.  The lexicographically first maximal independent set ($\LFMIS$ for short) is the lexicographically first maximal independent superset of the empty set.
To find the parent of~$S$, let $v$ be the vertex of  $\LFMIS\setminus S$ occurring earliest in the sequence,
let $N=N(v)\cap S$, and let $\parent(S)$ be the lexicographically first maximal independent superset of $(S\cup\{v\})\setminus N$. Each parent operation increases the length of the initial prefix of shared vertices between $S$ and $\LFMIS$, so after at most
$|\LFMIS|$ repetitions of the parent operation, the canonical maximal independent set
$\LFMIS$ will be reached.  Therefore, this parent operation satisfies the requirements of the reverse search paradigm.

For any vertex $v$ in $\LFMIS$, let $\later(v)$ be the set of vertices in
$V(G)\setminus\LFMIS$ for which $v$ is the adjacent $\LFMIS$ vertex
that appears earliest in the ordering.
The sets $\later(v)$ partition
$V(G)\setminus\LFMIS$.  If $S$ is maximal independent,
and $\parent(S)$ is the lexicographically first maximal independent
superset of $(S\cup\{v\})\setminus N$, then $N\subset\later(v)$,
for no vertex in $N$ could be adjacent to an earlier vertex in $\LFMIS$ than $v$
without violating the assumed independence of $S$.
Thus, to search for the children of a maximal independent set $S$, we need only consider sets $N\subset\later(v)$ for some $v$ in the initial prefix of shared vertices between $S$ and $\LFMIS$.

\begin{listing}[t]
\begin{lstlisting}
def search(S):
    output S
    for each vertex v in the ordered sequence:
        if v is not in LFMIS:
            continue
        if v is not in S:
            break
        for each nonempty independent subset N of later(v):
            T = (S union N) \ (neighbors of N)
            if T is maximal and parent(T) == S:
                search(T)
\end{lstlisting}
\caption{Recursive reverse search for maximal independent sets.}
\label{lst:recursive}
\end{listing}

\begin{listing}[t]
\begin{lstlisting}
def search(S):
    lseq = subsequence of vertices v with nonempty later(v)
    while True:
        output S
        v = first vertex in lseq
        N = first nonempty independent subset of later(v)
        while True:
            T = (S union N) \ (neighbors of N)
            if T is maximal and parent(T) == S:
                S = T   # unfolded call to search(T)
                break
            while True:
                if N != last nonempty independent subset of later(v):
                    N = next nonempty independent subset of later(v)
                    break
                if v != last in lseq:
                    v = next in lseq
                    if v in S:
                        N = first nonempty independent subset of later(v)
                        break
                if S == LFMIS:
                    return
                P = parent(S)
                v = first vertex from lseq in P \ S
                N = S \ P
                S = P   # unfolded return to search(P)
\end{lstlisting}
\caption{Nonrecursive reverse search for maximal independent sets.}
\label{lst:nonrecursive}
\end{listing}

The recursive search pseudocode in Listing~\ref{lst:recursive},
with an initial call to \lstinline$search(LFMIS)$,
instantiates the reverse search paradigm for the maximal independent set problem.
The collection of potential children generated by each call to \lstinline$search$
may include sets that are not maximal independent, so we must check that each generated set $T$ is maximal as well as that its parent is $S$; this problem of testing maximality is a large part of the difficulty in making our reverse search algorithms efficient.
To reduce the space occupied by the call stack, we generally prefer a version of the search
procedure in which the recursive calls have been unfolded into the procedure itself.
This unfolding can be done without need for an auxiliary stack, at the expense of some code complexity and an additional
$\parent$ computation per generated set,
as shown in Listing~\ref{lst:nonrecursive}.

\section{Minor-Closed Domination}

To make our search procedure efficient, we need a fast way of handling its maximality tests.  The sets that must be tested are automatically independent, and an independent set is maximal if and only if it {\em dominates} the graph; that is, all vertices are either in or adjacent to the set.  Therefore, we make use of a more general data structure for testing domination.  We describe here such a data structure for minor-closed graph families.
That is, given graph $G$ from a minor-closed graph family $\cal F$,
we wish to maintain a set $S\subset V(G)$, subset to insertions and deletions of
elements of $S$, and answer queries requesting the number of
vertices in $G$ that are not dominated by $S$.
We show how to do this in constant time per update or
query, and linear space and preprocessing time.
We will use this data structure later as part of our maximal
independent set enumeration algorithm.

\def\sameneighbors#1{\mathrel{\mathop{\sim}\limits^{\scriptscriptstyle #1}}}
\def\sameoutneighbors#1{\mathrel{\mathop{\sim}\limits^{\scriptscriptstyle #1+}}}

\begin{lemma}
\label{lem:few-supervertices}
For any minor closed graph family $\cal F$, graph $G\in\cal F$, and nonempty
$Q\subset V(G)$, define an equivalence relation
$\sameneighbors{Q}$ on $V(G)\setminus Q$ by $u\sameneighbors{Q} v$ if and only if
$N(u)\cap Q=N(v)\cap Q$. Then
the number of equivalence classes of $\sameneighbors{Q}$ is $O(|Q|)$,
where the constant of proportionality depends on $\cal F$ but not on $G$.
\end{lemma}

\begin{proof}
Suppose for a contradiction that we can find $G\in\cal F$ and $Q\subset V(G)$
where the number of equivalence classes of $\sameneighbors{Q}$ is an arbitrarily large
multiple of $|Q|$.  For any such $Q$ and $G$, form a minor $G_Q$ of $G$ by the following process:
let the vertices of $G_Q$ be the same as the vertices of $Q$.
Consider (sequentially, in arbitrary order) the equivalence classes of $\sameneighbors{Q}$,
choose a representative vertex $v$ for each equivalence class,
and form an edge in $G_Q$ by contracting a path through $v$ between
two vertices in $N(v)\cap Q$, unless all such paths connect pairs of vertices that are already
connected in $G_Q$.
At each step of the construction process outlined above, $G_Q$ is a minor of $G$;
therefore, it belongs to $\cal F$, and using the known sparsity of minor-closed graph families~\cite{Mad-MA-67} we can show that there are $O(|Q|)$ cliques in the partially constructed minor $G_Q$~\cite{ChiNis-SJC-85}.
An equivalence class represented by $v$ can only fail to add an edge to $G_Q$ iif
$N(v)\cap Q$ already forms a clique, so only $O(|Q|)$ equivalence classes fail to contribute an edge.
Thus, if the number of equivalence classes could be an arbitrarily large factor times $|Q|$,
we could form arbitrarily dense minors $G_Q$, contradicting the known bounds on
density of minor-closed graph families~\cite{Mad-MA-67}.
\end{proof}

One can prove more directly that planar graphs have at most
$\max(6|Q|-9,2|Q|)$ classes.

Our data structure consists of the vertices of $G$ together with a linear number of additional {\em supervertices}.  To form the supervertices, we form a sequence of graphs $G_0=G$, $G_1$, $G_2$, \ldots, as follows.  Let $\Delta$ be a constant, depending on graph family $\cal F$ but not on the particular graph $G$.  To form $G_i$ from $G_{i-1}$, let $Q_i$ be the set of vertices
in $G_{i-1}$ with degree at least $\Delta$.  Partition $G_{i-1}\setminus Q_i$ into the equivalence classes of $\sameneighbors{Q_i}$ and form a single supervertex for each such class.
Form the edges of $G_i$ from the induced subgraph for $Q_i$, together with
an edge from a supervertex $s$ to $h\in Q_i$ whenever a vertex $v$ in the equivalence class
corresponding to $s$ is connected to $h$ in $G_{i-1}$.

We can make the following observations and further definitions about this process.
\begin{itemize}
\item Each $G_i$ is isomorphic to a subgraph of $G$ (formed by choosing a representative vertex for each supervertex and omitting edges between pairs of supervertices), so it belongs to $\cal F$. 
\item By Lemma~\ref{lem:few-supervertices}, $|V(G_i)|=O(|Q_i|)$, and by choosing $\Delta$ sufficiently large
we can make $|Q_i\le\epsilon|V(G_{i-1}|$ for any constant $\epsilon>0$.  Thus, we can ensure that each graph in the sequence is smaller by a constant factor than the previous one.
\item Each supervertex has degree at most $\Delta-1$, so each vertex in $Q_i$ is an original vertex of $G$.
\item The sequence of graphs $G_i$ terminates only when the remaining graph forms a single supervertex of degree zero.
\item Each equivalence class of $\sameneighbors{Q_i}$ contains at most one supervertex of $G_{i-1}$ with the same degree.
\\item If a supervertex $v$ of $G_i$ contains a supervertex $u$ of $G_{i-1}$, and has the same degree, we consider the two to have the same identity, and represent them by the same object in the data structure.  
\item Define the {\em level} of a vertex or supervertex $v$ to be the largest $i$ such that $v\in V(G_i)$.
Each original vertex of $G$ belongs to at most $\Delta-1$ supervertices, which (if ordered by level) have a decreasing sequence of degrees.
\end{itemize}

\def\sv{\mathop{\rm sv}}
\def\nadj{\mathop{\rm nadj}}
\def\undom{\mathop{\rm undom}}
\def\nbr{\mathop{\rm nbr}}

\noindent
Our data structure consists of the following information.
\begin{itemize}
\item The graph $G$ and set $S$ to be maintained, and the set of vertices and supervertices in all graphs $G_i$ constructed as described above.
\item For each vertex or supervertex $v$ of level $i$, other than the degree-zero supervertex, a pointer $\sv(v)$ to the
supervertex corresponding to the equivalence class $v$ belongs to in $G_{i+1}$.
\item For each vertex or supervertex $v$ of level $i$, a count $\nadj(v)$ of the number of adjacent original vertices of $G$ that belong to $S$ and have level at most~$i$.
\item For each vertex of $G$ with level $i$, a list $\nbr(v)$ of adjacent vertices and supervertices in~$G_i$.
\item For each vertex or supervertex, a number $\undom(v)$.
For a vertex of $G$, $\undom(v)=1$ when $v\in V(G)\setminus S$
and $\undom(v)=0$ when $v\in S$.
For a supervertex,
$$\undom(v)=\sum_{\{w\,\mid\,\sv(w)=v \,\wedge\, \nadj(w)=0\}} \undom(w).$$
\end{itemize}

\begin{theorem}
\label{thm:minor-dom}
For any minor-closed graph family $\cal F$ and graph $G\in\cal F$,
the data structure above requires $O(n)$ space and can be constructed in $O(n)$ time.  We can query the number of undominated vertices in $G$ in $O(1)$ time, and insert or delete vertices in $S$ in $O(1)$ time per update.  All constants of proportionality in these bounds depend only on $\cal F$ and not on $G$.
\end{theorem}

\begin{proof}
The only nontrivial step in the construction of each graph $G_i$ is finding the equivalence classes of $\sameneighbors{Q_i}$, which can be done by bucket sorting in $O(n)$ time.  The times for constructing the whole sequence of graphs $G_i$ add in a geometric series to $O(n)$.

To query the number of undominated vertices, return $\undom(z)$ where $z$ is the degree-zero supervertex.  We say that supervertex $v$ is {\em reachable} from vertex $u$ if $u==v$ or $v$ is reachable from $\sv(u)$; $z$ is reachable from every vertex.
If $v$ is undominated, $\nadj(w)=0$ for each $w$ reachable from $v$, and $v$ contributes one to each reachable supervertex.  If $v$ is in $S$, it contributes zero to each reachable supervertex. If $v$ is not in $S$, but is dominated by a neighbor $u\in S$, let $w$ be the supervertex containing $v$ at the same level as $u$; then $\nadj(w)>0$ and $v$ does not contribute to any supervertices reachable from $w$.
Therefore, $\undom(z)$ is the number of undominated vertices in $G$.

To insert a vertex $v$ to $S$, increment $\nadj(w)$ for every $w$ in $\nbr(v)$,
and (if this causes $\nadj(w)$ to change from $0$ to nonzero) update $\undom(x)$ for every $x$ reachable from $w$.  Also update $\undom(v)$ and $\undom(u)$ for every $v$ reachable from $v$.
Each update takes constant time and a constant number of updates are performed, so the total time is constant.  Deletions are handled similarly.
\end{proof}

\section{Sparse Domination}

We now consider domination data structures for more general sparse graph families.
We say that graph $G$ is {\em $k$-orientable} if we can orient the edges of $G$
in such a way that each vertex has out-degree at most $k$.  Equivalently (by Hall's theorem), this condition asserts that every subgraph $H\subset G$ has at most $k|V(H)|$ edges.
Any subgraph-closed family of graphs with $O(n)$ edges per $n$-vertex graph is $k$-orientable for some constant~$k$; for instance, planar graphs are 3-orientable,
and a 3-orientation of a planar graph may be found in linear time~\cite{ChrEpp-TCS-91}.

Our data structure for domination in sparse graphs resembles that for minor-closed graph families, but differs in detail.
Given $k$-oriented graph $G$ and vertex set $Q\subset V(G)$, define equivalence relation
$\sameoutneighbors{Q}$ on $V(G)$ by $u\sameoutneighbors{Q} v$ if and only if
$N^+(u)\cap Q=N^+(v)\cap S$, where $N^+$ maps a vertex to its outgoing neighbors.
If $s$ is an equivalence class of $\sameoutneighbors{Q}$ containing~$v$, let $N^+(s)=N^+(v)\cap Q$.
We choose a (nonconstant) value $\Delta$,
let $Q$ be the set of vertices with degree at least $\Delta$, and create a supervertex for each equivalence class of $\sameoutneighbors{Q}$.
In our data structure we store the following data:

\def\lowdom{\mathop{\rm lowdom}}
\def\nundom{\mathop{\rm nundom}}
\def\hidom{\mathop{\rm hidom}}

\begin{itemize}
\item The dynamic set $S$ for which we wish to maintain dominance information.
\item For each vertex of $G$, the supervertex corresponding to its equivalence class.
\item For each vertex $v$ of $G$, the number $\lowdom(v)$ of edges $uv$ where $u$ belongs to $S$,
and where either $uv$ is oriented from $u$ to $v$ or $u$ has degree less than $\Delta$.
\item For each supervertex $s$, the number $\nundom(s)$ of vertices $v$ in its equivalence class which are not in $S$ and for which $\lowdom(v)=0$.
\item For each supervertex $s$, the number $\hidom(s)=|N^+(s)\cap S|$.
\item The sum of $\nundom(s)$, summed over those supervertices for which $\hidom(s)=0$.
\end{itemize}

\begin{theorem}
\label{thm:sparse-dom}
For any $k$-orientable graph $G$,
the data structure above requires $O(n)$ space and can be constructed in polynomial time.  We can query the number of undominated vertices in $G$ in $O(1)$ time, and insert or delete vertices in $S$ in $O(n^{1-1/k})$ time per update.
\end{theorem}

\begin{proof}
The query answer is given by the overall sum of $\nundom(s)$.

To insert vertex $v$ into $S$, increment the counts $\lowdom(v)$ of all outgoing neighbors of $v$, 
and (if $v$ has low degree) all incoming neighbors of $v$.  Update $\nundom(s)$ for each supervertex $s$ that has an equivalence class containing $v$ or one of these neighbors.
If $v$ has high degree, increment $\hidom(s)$ for all supervertices $s$ with $v\in N^+(s)$.  Whenever we change $\nundom(s)$ or $\hidom(s)$ we update
the overall sum.  The process of deleting a vertex is similar.

Inserting or deleting a low degree vertex takes time $O(\Delta)$, and inserting or deleting
a high degree vertex takes time $O((n/\Delta)^{k-1})$ since there are that many supervertices
associated with a fixed high degree vertex.
By choosing $\Delta=n^{1-1/k}$ we achieve the stated bounds.
\end{proof}

Dominating sets in $k$-oriented graphs can be used to model other problems including independent sets, matching, $k$-SAT, and constraint satisfaction.  As an example we show how to test independence in general graphs.

\begin{corollary}
\label{cor:dyn-ind}
We can maintain a set $S$ of vertices in an arbitrary $m$-edge graph $G$,
and test the independence of $S$, by a dynamic data structure that takes
time $O(1)$ per test and $O(\sqrt m)$ per insertion or deletion in $S$,
and uses linear space and preprocessing time.
\end{corollary}

\begin{proof}
Form a graph $G'$ having vertices corresponding to sets of 0, 1, or 2 vertices in $G$;
we include as 2-vertex sets in $G'$ only the sets of endpoints of edges in $G$.
Connect two vertices in $G'$ by an edge whenever the corresponding sets differ by a single element.
Then $G'$ can be 2-oriented by orienting all edges from larger sets to smaller sets.
The subset $S$ is independent in $G$ if and only if 
$\{\emptyset\}\cup\{\{v\}:v\notin S\}$
dominates all vertices of $G'$, so independence in $G$ can be tested by our domination algorithm for the 2-oriented graph $G'$.
\end{proof}

The same result can be achieved more directly:
In any $k$-oriented graph, we can maintain the number of adjacent pairs in a dynamic set $S$, and therefore determine the independence of $S$, by a simple data structure that stores for each vertex the number of incoming edges from vertices in $S$, in time $O(k)$ per update.  The corollary follows since any $m$-edge graph can be $O(\sqrt m)$-oriented.
However, the proof we have given for Corollary~\ref{cor:dyn-ind} provides some evidence that dominance is strictly harder to maintain than independence, since any improvement to Theorem~\ref{thm:sparse-dom} for $2$-orientable graphs would lead to a corresponding improvement to Corollary~\ref{cor:dyn-ind} for arbitrary graphs.

\section{Sparse Independent Sets}

A graph $G$  is {\em $k$-degenerate}~\cite{MatBec-JACM-83,SzeWil-JCT-68} if its vertices
can be ordered in such a way that, for each vertex, the number of neighbors occurring later in the ordering is at most~$k$.  Equivalently, each subgraph of $G$ has a vertex with degree at most~$k$. This parameter is also known as the inductiveness or the Szekeres-Wilf number of $G$.  A $k$-degenerate ordering of $G$, if one exists, can be found by a simple greedy algorithm
in linear time~\cite{MatBec-JACM-83}.  It is known~\cite{Mad-MA-67} that all minor-free graph classes have bounded degeneracy.  A $k$-degenerate graph is clearly $k$-orientable (simply orient each edge from the earlier to the later vertex in a $k$-degenerate ordering) and conversely
a $k$-orientable graph must be at most $2k$-degenerate, so graphs of bounded degeneracy are the same as the graphs of bounded orientability considered in the previous section.

\begin{theorem}
Let $G$ be a $k$-degenerate graph in which we can maintain a dynamic set $S$ and test whether $S$ dominates $V(G)$ in time $T(n)$ per insertion or deletion to $S$.  Then, we can list all
maximal independent sets in $G$, in time $O(nT(n))$ per generated set and polynomial delay.
The space required by the algorithm is $O(n)$ plus a single instance of the dynamic dominance data structure.
\end{theorem}

\begin{proof}
We use the $k$-degenerate ordering as the vertex ordering for our reverse search algorithm.  Therefore, each set $\later(v)$ will have
at most $k$ vertices, each vertex in $G$ will participate in $O(1)$ sets $N\subset\later(v)$ throughout a call to search,
and the sum of the numbers of neighbors of sets $N$ will be proportional to the number of edges
in the graph, which is $O(n)$.  To save space, we unfold the recursive calls of the search into a nonrecursive version of the algorithm,
as described in Listing~\ref{lst:nonrecursive}.  The delay bound follows since we can at most test $O(n)$ potential children each at $O(n)$ levels of the recursion before outputting another set or exiting the search.

As the algorithm progresses, we maintain a dynamic dominance data structure for the current set~$S$, which we use to test each successive set
$T$ for maximality with a number of updates proportional to the size of the set of neighbors of $N$; thus, throughout a call to the recursive version of the search algorithm, the number of data structure updates is $O(n)$.  In the nonrecursive search, we also update the data structure
whenever we change the set $S$ by an unfolded recursive call or return;
the total number of data structure updates caused by these changes is again $O(n)$ per output set.

Finally, we must consider the time taken to compute $\parent(T)$ for each
potential child $T$ considered by the search algorithm.  In these computations,
$\parent(T)$ is the lexicographic maximal independent superset of the independent set
$(T\cup\{v\})\setminus N$.  We also know (from the maximality of $T$) that the only
vertices that can be added in forming the lexicographic maximal independent superset are
neighbors of $N$.  To perform this computation efficiently, we maintain a simple data structure
that stores a count in each vertex of the number of incoming edges from vertices in $S$.
Each change to $S$ causes $k$ counts to be updated, in time $O(1)$.
Then, we modify this data structure to count incoming edges from $T$ instead of $S$,
and use it to compute the lexicographic maximal independent superset of $T$
in time proportional only to the number of neighbors of~$N$.  Therefore, all parent computations can be done in time $O(n)$ per generated set.
\end{proof}

\begin{corollary}
Let $\cal F$ be a minor-closed graph family, and let $G$ be an $n$-vertex graph in $\cal F$.  Then we can list all maximal independent sets in $G$, in time $O(n)$ per set, space $O(n)$, and polynomial delay, where the constants of proportionality depend on $\cal F$ but not on $G$.
\end{corollary}

\begin{corollary}
We can list all maximal independent sets in any $n$-vertex $k$-oriented graph in time $O(n^{2-1/k})$ per set,
space $O(n)$, and polynomial delay.
\end{corollary}

\section{Bounded Degree Independent Sets}

We now briefly describe our algorithm for bounded degree graphs.
The key observation in this case is that, for a given maximal independent set $S$, vertex $v\in S\cap\LFMIS$, and $N\subset\later(v)$, the associated set $T$ is a child of $S$ if and only
if the following three conditions hold:
\begin{enumerate}
\item $T$ is maximal.
\item $S$ is the lexicographically first maximal independent superset of $(T\cup\{v\})\setminus N$.
\item Vertex $v$ is the earliest vertex of the sequence in $\LFMIS\setminus T$.
\end{enumerate}

The first two conditions depend only on the inclusion or exclusion in $S$ of a constant number of
vertices within distance $O(1)$ of $v$.  In particular, $T$ is non-maximal if and only if some vertex within
distance three of $v$ can be added to $T$, which can be tested by examining all vertices within distance four of $v$.  If $T$ is maximal, the computation of the lexicographically first maximal independent superset of $(T\cup\{v\})\setminus N$ can be done by examining only vertices within distance two of $v$.

We say that a pair $(v,N)$ is {\em fertile} for a set $S$ if the set $T$ generated from $S$ using $v$ and $N$ passes the first two of the three conditions listed above.  The third condition can be rephrased as stating that $v$ belongs to the initial common prefix of $\LFMIS$ and $S$.  As our algorithm progresses,
we maintain the following information:
\begin{itemize}
\item The position of the last vertex in the initial common prefix of $\LFMIS$ and $S$.
\item The set of all fertile pairs $(v,N)$ where $v$ occurs before the last initial common vertex,
stored as a dictionary mapping each vertex $v$ to the sets $N$ that form fertile pairs for it.
\end{itemize}

The pairs in the set maintained by the algorithm give exactly the children of the current maximal independent set $S$.  Each child differs from $S$ in a constant number of vertices, and so
the set of fertile pairs for the child also differs by a constant.  However, the position of the last vertex can differ dramatically between $S$ and its children.  In order to keep the changes to the set of fertile pairs gradual, we modify our search algorithm so that it processes the children of $S$ in the reverse of the vertex ordering on the vertices $v$.
 
\begin{listing}[t]
\begin{lstlisting}
def search(S, last_common, fertile_pairs):
    output S
    for v in fertile_pairs, in reverse order by vertex sequence:
        later_subsets = fertile_pairs[v]
        delete v from fertile_pairs
        last_common = predecessor of v in LFMIS
        for N in later_subsets:
            S = (S union N) \ (neighbors of N)
            update fertile_pairs from changes to S
            search(S, last_common, fertile_pairs)
            S = LFMISS((S union {v}) \ N)
            update fertile_pairs from changes to S

initial call:
    S = LFMIS
    last_common = last vertex in S
    F = fertile pairs for S
    search(S, last_common, F)
\end{lstlisting}
\caption{Reverse search for maximal independent sets in bounded degree graphs.}
\label{lst:bounded-deg}
\end{listing}

Thus, the maximal independent sets in $G$ can be generated by the algorithm described
in Listing~\ref{lst:bounded-deg}.  Each time we change $S$ to form one of its children,
we can potentially affect fertile pairs for vertices within distance four of the change;
for each such vertex $w$ occurring no later than \lstinline$last_common$ in the vertex sequence, and each independent subset $N$ of $\later(w)$, we test whether the change to $S$ has caused $(w,N)$ to start or stop being a fertile pair, and if so add or remove $N$ to or from \lstinline$fertile_pairs[w]$.
There are $O(1)$ pairs $(w,N)$ tested per change to $S$, so the total time per child is $O(1)$.

We have not described what order to use for the vertices, because any ordering will work for the correctness of the algorithm and its asymptotic analysis.  However, a $k$-degenerate ordering for the minimum possible~$k$ may be preferable to other orderings, because it reduces the number of subsets of $\later(v)$ that need to be considered for each $v$ and thereby reduces the constant factors in our analysis.

One complication with the analysis of the algorithm above is the question of how we maintain or sort the vertices $v$ considered by the outer loop, so that they are considered in reverse order.
For this analysis, we assume a simple comparison sorting algorithm that sorts these vertices in time
$O(k\log k)$, where $k$ is the number of vertices to be sorted.

\begin{lemma}
\label{lem:bd-recharge-targets}
If there are $k$ vertices in the {\rm\lstinline$fertile_pairs$} data structure for maximal independent set $S$, then $\Omega(k)$ of the children of $S$ have $\Omega(k)$ vertices in their respective
{\rm\lstinline$fertile_pairs$} data structures.
\end{lemma}

\begin{proof}
The changes in the maximal independent set from $S$ to its children lead to $O(1)$ changes to the \lstinline$fertile_pairs[w]$ data structure, and each time we consider a vertex $v$ in the outer loop for $S$ we remove only that vertex from the data structure.  So, for each of the first $k/2$ vertices considered, there remain $k/2-O(1)$ vertices in the data structure for the children.
\end{proof}

\begin{theorem}
Let $G$ be a graph with maximum vertex degree $O(1)$.  Then, we can list all
maximal independent sets in $G$, in time $O(1)$ per generated set, space $O(n)$, and polynomial delay.
\end{theorem}

\begin{proof}
We use the recursive version of the algorithm described in Listing~\ref{lst:bounded-deg}, which modifies the set $S$ and the data structure \lstinline$fertile_pairs[w]$ in-place and shares the modified structures with each recursive call.  The space for these structures is $O(n)$, the other space used by the algorithm per call is $O(1)$, and the call stack may be $O(n)$ levels deep, so the total space is $O(n)$.  Polynomial delay follows as before.  As we have seen, the time per child is $O(1)$, except for the time spent sorting the vertices in the \lstinline$fertile_pairs[w]$ data structure prior to looping over them.
That time is $O(k\log k)$, and we charge the time spent in this step equally to each of the $\Omega(k)$ children described in Lemma~\ref{lem:bd-recharge-targets}.  In this way, each recursive call gets
charged $O(\log k)$ time, negligible compared to the $O(k)$ time the recursive call will spend on listing its own children.
\end{proof}

\section{Conclusions}

We have provided a general reverse search based framework for generation of all maximal independent sets, and applied it to many important graph classes.  Along the way we were led to study new dynamic graph data structures for independence and domination.

One natural problem for additional research is to quantify and reduce the dependence of the running time on the sparseness of the graphs in question.  Our maximal set generation algorithms depend in an exponential way on the sparseness parameter (orientability, degeneracy, or degree) of the graphs we consider, due to the way we examine all independent subsets of the sets $\later(v)$.  Can this exponential dependence be reduced?

Also, to what other non-sparse graph classes can our results be extended?  The class of graphs that can be ordered so that $|\later(v)|=O(1)$ for all $v$ contains non-sparse graphs, but it is not clear how to perform the dominance or parent tests needed by our algorithms efficiently in such graphs.  However,
we were able to extend our technique to some other graph classes where it did not improve previous results.  For chordal graphs, if we use
an elimination ordering, each independent subset $N\subset\later(v)$ has exactly one vertex,
and (by techniques involving counting uniquely dominated vertices) we were able to achieve $O(m)$ time per independent set, matching a prior result of Leung~\cite{Leu-Algs-84}.
Maximal independent sets for interval graphs may be translated to paths on an associated digraph (connect intervals I and J by an edge if J is to the left of I and no interval K is between the two), so all maximal independent sets or more general weighted $k$-best independent set generation problems can be solved in constant time per set by a $k$-shortest paths algorithm~\cite{Epp-SJC-98} unrelated to the present reverse search approach; a similar approach also works for generating chains and antichains in two-dimensional dominance relations.  For intersection graphs of disks, balls, squares, or cubes, if we order the objects by size then each independent subset of $\later(v)$ must have a constant number of objects, so our algorithm can be made to perform in polynomial time per output set, but the polynomial appears larger than for general graph independent set enumeration algorithms.

\raggedright
\bibliographystyle{abuser}
\bibliography{misenum}
\end{document}